# HiFlash: A History Independent Flash Device

Bo Chen
College of Information Sciences and Technology
The Pennsylvania State University
bxc30@ist.psu.edu

Radu Sion
Department of Computer Science
Stony Brook University
sion@cs.stonybrook.edu

*Abstract*—Retention regulations require timely and irrecoverable disposal of data, a challenging task, as data and its side effects are stored and maintained at all layers of a computing system. Those side effects can be used as an oracle to derive the past existence of deleted data.

Fortunately, history independence can be utilized to eliminate such history-related oracles. HIFS [19] can provide history independence for file storage over mechanical disk drives. However, HIFS cannot provide history independence when deployed on top of flash devices, as flash memory manages its own internal block placement, which is often inherently history dependent.

In this work, we initiate research on history independent flash devices. We design HiFlash, which achieves a strong notion of history independence by defending against an adversary allowed access to the flash at multiple different points in time. In addition, we design a simple, yet history independence friendly wear-leveling mechanism that allows HiFlash to smartly and advantageously trade off a tunable small amount of history leakage for a significant increase in the device's lifetime. Our prototype built in an actual flash device as well as extensive simulations validate the effectiveness of HiFlash.

## I. INTRODUCTION

Regulations such as the Health Insurance Portability and Accountability Act (HIPAA) [25], the Gramm-Leach-Bliley Act [17], and the Sarbanes-Oxley Act (SOX) [51], mandate consistent procedures for information access, processing, and storage in health care, financial services, public corporations and government sectors. Regulated organizations are liable for the improper management of data even after its legal expiration. Thus, of paramount importance are data life-cycle regulations, that define the process of removing data once it legally expires.

Ensuring complete irrecoverability of deleted data is difficult to achieve in modern systems. Simply overwriting data or deploying encryption with ephemeral keys is not sufficient since the write history itself implicitly leaves artifacts in the layout of the resulting storage medium at all layers. The artifacts can then be used as an oracle to answer questions about the past existence of deleted records.

For example, the current layout of data blocks on disk is a direct and often fully deterministic function of the sequence and timing of previous writes to file system, database search indexes, etc.

Questions such as "was Johns record ever in the HIV patients dataset" can then be answered much more accurately than guessing by simply looking at the storage layout of the search index on disk – which will look different (e.g., with a 30% likelihood) depending on whether John has previously been in the data set or not.

These are the very questions that secure deletion promises to prevent anyone (including insiders) from answering once Johns record has been deleted.

And, unfortunately, in this case, the security of a potential secure deletion mechanism is reduced from an apparently "strong" 256 bit encryption with ephemeral keys to a ONE IN THREE chance (30%) of determining whether John had HIV!

In other words, the mere (previous) existence of deleted records impacts the current system state implicitly at all layers. This can be used as an oracle to derive information about the past existence of deleted records. However, if all system layers are designed to exhibit history independence, such implicit history-related oracles can be made to disappear.

To enforce data life-cycle regulation requirements, it is imperative to conceal historical information contained within data structure states. This can be achieved by using history independent data structures [32, 46] to organize the data. A data structure is said to be history independent if its current layout is not impacted by its history. By applying history independent data structures to storage organization, an attacker, when accessing the current layout of the storage, cannot learn the history of the past operations.

Prior work focused on designing various history independent data structures [20, 30, 31, 46, 54]. Little effort has been made to tackle challenges in deploying history independent data structures in systems. Nevertheless, achieving history independence efficiently is hard due to the fact that current systems are designed to heavily benefit from (data and time) locality at all layers through heavy caching, and existing history independent data structures completely destroy locality.

Bajaj et al. [19] designed HIFS, which is till now the sole effort of building history independent systems. HIFS however, can only provide history independence for file storage over mechanical disk drives. When the underlying storage media are changed to flash devices (e.g., SSDs), HIFS cannot provide history independence due to the following reasons:

(a) Flash memory has a limited number of program-erase (P/E) cycles. To avoid being worn out soon, a flash device usually manages its own internal block placement, which may lead to history breaches, because such "wear leveling" placement is often inherently history dependent. Firstly, to avoid repeatedly writing the same flash cell, the flash device usually allocates new space for a write, regardless if it is a new write or an over-write. These allocators in the commodity flash devices are designed without considering history independence and may not preserve history. For example, a linear allocator allocates free space to the new writes sequentially, which directly conflicts with history independence as the resulting layout depends on the order of writes. Secondly, a flash device usually performs wear leveling to balance erasures across flash cells. This wear leveling mechanism usually requires to periodically relocate blocks according to the write history, which will compromise history independence.

(b) A flash device does not guarantee instantaneous deletion. Due to its limited P/E cycles, it usually introduces deletion latencies in order to reduce the number of block erasure operations. In the interval between an application-issued erase command and the time when the flash device actually erases the block, the erased data can be recovered by an attacker with access to the device.

Flash devices have been used extensively in mobile devices like tablets and smart phones. Even in PCs, flash devices under the form of SSDs have gained popularity nowadays. According to Gartner [11], the global SSD market size had reached $10.9 billion in 2013. Thus, providing history independence for flash devices is of great importance, was posed as an open problem [19] and unfortunately is still unsolved.

To provide history independence for flash devices, we need to handle several challenges: Firstly, the allocators used in the existing flash devices are not history independent. Designing a history independent allocator for flash memory is challenging; Secondly, as an essential functionality of flash controllers, wear leveling inherently needs to interfere with history independence, since it may often need to re-locate blocks according to write history. Wear leveling without compromising history is thus highly non-trivial. Thirdly, history independence may require to perform in-place updates, which are expensive to achieve in flash and may lead to significant write amplification, because over-writing a small portion of data in flash may require erasing and rewriting a large portion of data. How to accommodate in-place updates and mitigate write amplification is not straightforward.

In this work, we address the aforementioned challenges and design HiFlash, the first History independence schemes for Flash-based block devices. HiFlash is an essential component that can be leveraged in a multi-layer approach to provide history independence at the block device layer for SSDs, the dominant high-end storage medium. Our contributions are summarized as follows:

(1) We initiate research on history independence for flash-based block devices.
(2) We design HiFlash, a first solution providing history independence for flash-based block devices. HiFlash achieves a strong notion of history independence by defending against an adversary allowed access to the flash at multiple different points in time.
(3) We optimize HiFlash by mitigating write-amplification. We design a history independence friendly wear leveling mechanism that allows HiFlash to smartly and advantageously trade off a tunable small amount of history leakage for a significant increase in the device's lifetime.
(4) We implement HiFlash in an actual flash device using the OpenNFM framework.
(5) We measure the performance and impact on the underlying device. Results are encouraging. Sequential read throughput is within $0.9\times$ and random read throughput is within $1.2\times$ of baseline OpenNFM; write operation performance is $0.2\times$ due to write amplification, a possibly unavoidable price to pay for history independence. Leakage quantification and simulations seem to confirm that epoch-based wear leveling balances a good leakage - wear leveling effectiveness trade-off.

## II. BACKGROUND

**History Independence.** History independence aims to prevent historic information about the pattern of access to a data structure from being leaked through its representation when observed by an external party. We consider two types of history independence [33, 46]: weak history independence (WHI) and strong history independence (SHI). WHI allows an adversary to observe the data structure a single time, whereas SHI allows multiple observations over time. A data structure implementation is said to be history independent if nothing can be learned from the data structure's memory representation during these observations except for the current abstract state of the data structure [33]. If a data structure has canonical representations for each state, it is necessarily SHI; conversely, SHI necessarily implies that the data structure has canonical representations up to initial randomness [33].

To provide history independence, we can use a history independent hash table [20, 46]. Similar to conventional hash tables, a history independent hash table uses hash functions to probe the table. However, the history independent hash table uses a different collision resolution mechanism (e.g., priority functions [46] or Gale-Shapley Stable Marriage [20]) that can guarantee the resulting layout is independent of the input patterns.

**Flash Memory.** Flash memory is a non-volatile computer storage medium which can be electrically erased and re-programmed. It can avoid the mechanical limitations of hard disk drives, and has significant advantages on speed, noise, power consumption, and reliability. There are two main types of flash memory, NAND flash and NOR flash. In this work, we focus on NAND flash, which is widely used in flash-based block devices like USB sticks, solid state drives (SSDs), MultiMediaCards (MMCs), SD cards. The NAND flash array is usually grouped into blocks, each of which is a collection of pages. A concrete organization of NAND flash is shown in Figure 1. Typically, a flash block contains 32, 64, or 128 pages. Each page can be 512, 2048, or 4096 byte. Associated with each page are a few bytes (i.e., the *spare area*), which is usually $\frac{1}{32}$ of the page size and can be used to store the error-correcting code (ECC). NAND flash has several known limitations: block erasure, memory wear, and read/program disturb.

(a) Block erasure: Flash memory must be erased before it can be re-written (i.e., *erase-then-write*). For NAND flash, reading and programming (writing) are performed on a page basis, but erasure can only be performed on a block basis (a flash block is also called "erase block" or "physical block"). This however, may lead to a situation in which the actual amount of information actually written to the flash is a multiple of the amount of information written by the host system. This is known as *write amplification*. In general, write amplification is the ratio between the data that ends up written to the flash and the data written by the host. To re-write a small portion of data in flash memory, we can have two options: 1) read all the data from its corresponding erase block, update this small portion of data, erase the entire block, and write all the data back; 2) read this portion of data from its corresponding erase block, update it, and write it to a new block (the new block, if previously containing data, would need to be erased, and the old data stored would need to be relocated before an erasure can be performed). In both cases, the data actually written to the flash memory is larger than the data written by the host.

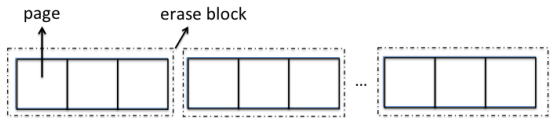

Fig. 1. The organization of NAND flash.

(b) Memory wear: NAND flash has a limited number of program-erase (P/E) cycles before the wear begins to deteriorate the integrity of the storage (i.e., blocks are compromised and cannot hold data any longer). Typically, SLC (single-level cell) NAND flash rates around $100K$ P/E cycles, while MLC (multi-level cell) NAND flash rates around $1K - 10K$ P/E cycles [3]. To prolong the service life of flash memory, data should be placed such that erasures and re-writes are distributed evenly across it. The idea is to flatten out the access rate distribution and ensure that no single block is much hotter than the average and fails prematurely. This is known as *wear leveling*. Conventional wear leveling relies on erasure counts to swap blocks. Qureshi et al. [48] proposed start-gap wear leveling, which periodically moves data around flash blocks in a fixed pattern, and thus no need to keep track of erasure counts.

(c) Read/Program disturb: In NAND flash, memory cells (each cell holds a flash page) are serially connected in a string structure, and reading/programming a cell may cause its nearby cells in the same block to change over time. This is known as *read/program disturb*. This usually happens after hundreds of thousands of reads/programmings. Since NAND flash usually computes an error-correcting code (ECC) for each page, a read/program disturb error may be corrected until the number of its bits affected by read/program disturb exceeds the number of bits the ECC can recover.

**MTD (Memory Technology Device).** Flash is often exposed through standard abstraction layers such as MTD [9]. MTD hides many aspects specific to particular flash chips, and provides uniform APIs to access different types of flash memory. The uniform APIs provided by MTD are MTD_Read, MTD_Write, and MTD_Erase, which allow to perform read, write, and erase over raw flash respectively.

**UBI (Unsorted Block Images).** UBI is usually built on top of MTD devices. It manages logical volumes on a single physical flash device, and implements two main functionality, wear leveling and bad block management. UBI implements wear leveling such that continuous writes/erasures will be spread to all the flash blocks. In addition, UBI is aware of bad erase blocks, and transparently handles those bad blocks.

**Flash-specific File Systems and Flash Translation Layer.** To utilize a raw MTD flash device, we can either use a flash-specific file system or expose the flash memory as a block device: (1) A flash file system is a file system optimized specifically for flash memory. Popular flash file systems include JFFS2 [5], YAFFS [16], UBIFS [14], LogFS [6]. For example, unlike conventional file systems, UBIFS always picks different flash blocks for journal when the current journal is filled, to avoid moving data out of the journal. (2) A flash-based block device provides block device functionality to an external party. This allows a conventional file system (e.g., ext4, FAT32) to use yet be agnostic of the underlying flash device. Due to the specifics of flash, this naturally leads to suboptimal utilization and device life-cycle issues. In practice, this is achieved through a Flash Translation Layer (FTL), a built-in controller for flash devices. FTL translates logical block addresses to physical flash addresses, and exposes a block device interface. Most of the commercially available flash devices (e.g., SSDs, USB sticks, MMCs, SD cards) have a built-in FTL. **In this work, we mainly consider flash devices which are exposed as block devices by using FTL, e.g., SSDs.** Note that Traditional FTLs [35, 39] usually adopt a variant of log-structured writing mechanism. They may not be able to provide history independent flash layout, as their flash layout is usually a function of the writing history due to "log-structured writing".

## III. MODEL

### A. System Model

We consider a flash that consists of $m$ erase blocks, each composed of $l$ pages (Figure 1). Using FTL (Sec. II), flash can be exposed through a block-based access interface. In other words, the whole device is represented as a linear array of $N$ "virtual blocks", each of which may be read or written by an external party (e.g., an operating system). We call these "virtual blocks", in order to differentiate them from the flash erase blocks. Generally, we have $N * |virtual\ block| \leq m * l * |flash\ page|$, as the flash may need to reserve space for storing system meta-data, besides storing all the data being written to the $N$ virtual blocks. The interface for block-based provides the following entry-points: (let $i$ be the virtual block ID and $0 \leq i \leq N-1$):

- Block_Read($i$, &block): read data from virtual block $i$
- Block_Write($i$, block): write data to virtual block $i$

NAND flash usually has 512, 2048 or 4096 byte minimum input/output unit size (min I/O) [8], corresponding to the underlying flash page sizes, and writes to NAND flash need to be aligned to min I/O, and in increments being multiples thereof. Thus, for efficiency, the size of a virtual block should be a multiple of min I/O, i.e., a multiple of flash page size. **To simplify our presentation, we assume the size of a virtual block is equal to the size of a flash page. However, our solutions are easily extended to the case where the virtual block size is a multiple of the flash page size.** When data is written to a virtual block, it will be stored at a flash page.

**Flash Allocator.** The mechanism that assigns virtual block writes to actual flash pages is called *flash allocator*. Let F be a function used by the flash allocator to compute the location of the flash page for storing the data being written to a virtual block. F is defined as $\text{F}: \{0,1\}^{logN} \to \{0,1\}^{logm} \times \{0,1\}^{logl}$, where $N$ is the total number of virtual blocks, $m$ is the total number of erase blocks in flash and $l$ is the number of pages in an erase block. $\text{F}(i) = (x, y)$ means the data written to a virtual block $i$ will be stored at flash page $y$ of erase block $x$.

### B. Adversarial Model

We consider a *one-access adversary* and a *multiple-access adversary*. The one-access adversary can have access to the flash memory only once, e.g., by stealing a laptop or a smart phone. The multiple-access adversary can have multiple access to the flash memory, e.g., by periodically breaking into a hotel room and obtaining a "memory dump" of the targeted flash device. Both adversaries aim to illegitimately derive sensitive information which is not available through a "legitimate" interface, including the order of past operation sequence (e.g., the order in which ballots were cast in a voting machine [43]), the evidence of past existence of delete data [18], etc.

## IV. SCENARIOS

In the following, we provide scenarios showing that commodity SSDs may not provide history independence, and the adversary may take advantage of this to either infer the past existence of deleted data, or compromise the order of past operation sequences. Note that commodity SSDs usually prefer a log-structured writing technique [10], by which data and metadata are written sequentially to pages of a flash block. Let $A$, $B$, $C$ and $D$ be the data written to flash.

**Scenario 1: inferring the previous existence of deleted data.** As shown in Figure 2(a), $A$, $B$ and $C$ are written to flash in step 1. In step 2, to securely delete B (e.g., after a TRIM operation [13] is issued), the corresponding page can be cleared by using zero overwriting [52] or scrubbing [55]. In step 3, $D$ is written. As flash memory must be erased before it can be re-written (Sec. II), $D$ needs to be written to a new page, rather than the old page storing $B$ previously. By having access to the flash layout, an adversary may be able to tell there was a deleted record stored between $A$ and $C$. As $B$ may be correlated to both $A$ and $C$, the adversary may infer more information about the deleted record $B$.

**Scenario 2: compromising the order of past operation sequences.** As shown in Figure 2(b), for three different write sequences, the resulting layouts in flash are different. By having access to the flash layout, an adversary may be able to tell which write sequence leads to a certain state. For example, if the adversary finds out the current flash layout is of case 2, it will know that the write sequence was not "A, B, C" or "C, A, B", very likely "B, C, A". If $A$, $B$ and $C$ are voting records, the adversary may know the order in which people voted, leading to compromise of voter privacy [43].

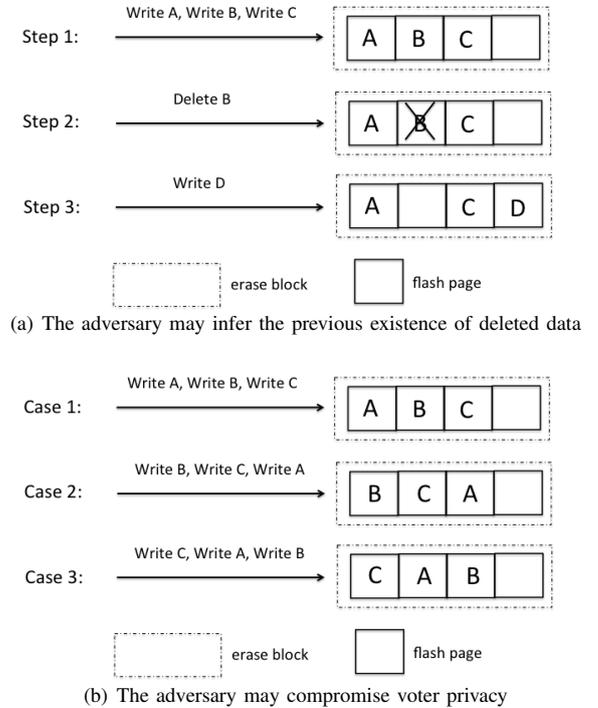

(a) The adversary may infer the previous existence of deleted data

(b) The adversary may compromise voter privacy

Fig. 2. Motivating scenarios.

## V. HiFlash

In this section, we present HiFlash, a collection of history independent schemes and associated implementations for flash-based block devices.

### A. Key Insights

Multiple key insights lead to our HiFlash designs.

**Using a history independent data structure without collision resolution.** Seemingly, we can directly use a history independent hash table (Sec. II) when designing the flash allocator. This would provide history independence yet would be impractical, for a number of reasons. The performance of a hash table will degrade significantly when its load factor is large [19], as collisions will become frequent, leading to frequent re-locations of data. This will be exacerbated in flash due to its erase-then-write requirement (Sec. II). In our setting (Sec. III-A), a hash table seems to be unnecessary. Originally designed to map a key from a large domain to a value stored in a small array, the hash table usually has additional overhead for collision resolution. A flash allocator maps the keys from a small domain (i.e., $[0, N-1]$, where $N$ is the total number of virtual blocks) to a value stored in a small array, which does not require using a hash table and may avoid the overhead of collision resolution.

**Temporal locality can mitigate write amplification.** A vast majority of workloads of interest exhibit a certain temporal locality property as our experiments and others have shown [34, 40]. Recently written blocks are likely to be written again in the near future. However, in flash memory, performing an in-place update on a page requires to first erase its entire encompassing erase block. This may lead to significant write amplification if the other pages of this block are not empty. We may be able to mitigate this by caching multiple subsequent writes and performing them together, as the cached writes may target only a few different erase blocks, and by waiting, we reduce the number of total block erasures necessary per incoming write.

**Wear leveling that balances a leakage – wear leveling effectiveness trade-off.** Wear leveling is of paramount importance for flash memory as each flash cell usually features a limited number of program-erase cycles (Sec. II). The rationale of wear leveling is to move hot data around, such that writes and erasures can be distributed evenly among flash cells, prolonging the service life of flash memory. However, to identify hot data, conventional wear leveling techniques [23] usually need to keep track of the entire write history, which may lead to significant history leakage. Our design used a controlled, significantly smaller amount of historical information to achieve effective wear leveling, achieving an acceptable leakage – wear leveling effectiveness trade-off.

### B. HiFlash *Schemes*

In the following, we first present a basic HiFlash scheme, which mitigates write amplification by leveraging the temporal locality of its input access patterns. We then present an improved HiFlash scheme, which adds an optimized history independence friendly wear leveling mechanism that can trade off a tunable small amount of history leakage for a significant increase in the device's lifetime.

*1) A Basic* HiFlash *Scheme:* This basic scheme is built around a bijective mapping between virtual block IDs and flash pages. A potential flash allocator for this setting is illustrated in Figure 3, in which F (Sec. III-A) is constructed around a simple permutation $\Pi$ on the set of virtual block IDs as follows: $F(i) = (\Pi(i)/l, \Pi(i)\%l)$, where $i$ is the virtual block ID.

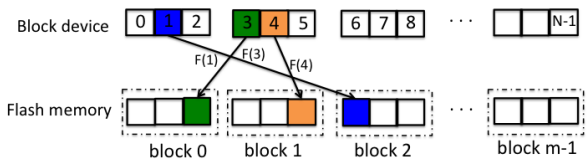

Fig. 3. A history independent flash allocator.

**Mitigating write amplification.** When an external party overwrites a virtual block on the block device, our history independent flash allocator needs to perform an in-place update over the corresponding flash page. This may cause write amplification (Sec. II) because, it may require an erase of the corresponding erase block, which may lead to multiple additional page reads/writes to accommodate any of its non-empty pages.

One idea to mitigate this may be to increase the size of the virtual block. At one extreme, the virtual block size can be equal to the size of an erase block, in which case, seemingly, write amplification resulted from in-place updates will be completely eliminated. However, a number of representative workload access pattern data sets suggest that in many scenarios, most of the writes performed on a block device are only a few KBs in size (see Sec. VII-B). Thus, choosing the virtual block as large as an erase block may end up being extremely inefficient in practice and in fact lead to an even higher overall device degradation over time.

Another idea would be to use the temporal locality of incoming writes (Sec. V-A). Firstly, data being written to adjacent virtual blocks will be more likely stored in the same erase block. We call this new property *locality preservation*. A locality-preserving history independent flash allocator is shown in Figure 4, $F(i) = (x, y)$, where $x = \Pi(i)/l$, $y = \Pi(i)\%l$ and $\Pi(i) = i$. The locality-preserving history independent flash allocator is also a bijective mapping between virtual block IDs and flash pages.

Secondly, a certain number of writes can be cached and performed together, to preserve a certain degree of locality in flash. The number of erasures required will be reduced as the cached writes will necessarily belong to fewer different erase blocks.

One security concern here is the history leakage associated with cached writes, which may now be available to an adversary getting access to the device before the writes have been flushed. Using a history independent data structure [28] to organize the cached writes helps mitigate the leakage, but unavoidably, the attacker will know that the writes being cached are recent.

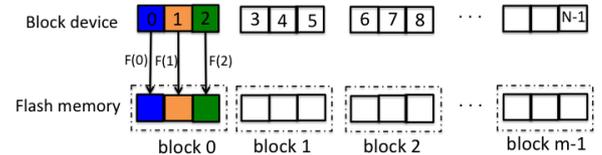

Fig. 4. A locality-preserving history independent flash allocator.

*2) An Improved* HiFlash *Scheme:* The basic HiFlash scheme uses a history independent flash allocator, bijectively mapping virtual blocks to flash pages. However, conventional file systems (e.g., FAT, NTFS and ext2) were originally designed for magnetic disks, and may re-write many of their data structures repeatedly to the same area of a block device. The basic HiFlash scheme does nothing to handle any such distribution unevenness – some flash regions will end up significantly "hotter" than others and risk early wear and failure. This is confirmed by real workload data (see Sec. VII-B). To prevent premature device demise, we need to provide wear leveling.

At its core, any wear leveling mechanism basically swaps the location of hot and cold pages in the hope of "evening out" overall wear. Naturally, swapping itself features a write overhead and thus needs to be performed smartly lest it defeats the very purpose it was designed for. Conventional *global wear leveling* [23, 24] usually keeps track of lifetime block or page "erasure counts", to decide which blocks or pages to swap and where to place them. This can help achieve optimal wear leveling [23]. However, lifetime erasure counts may leak an unacceptable amount of history and result in a history-dependent layout.

**Strawman solution: random wear leveling.** To start mitigating this issue, we first propose a simple wear leveling solution without the above leaks: periodically choose a number of blocks uniformly at random and swap them. We call this solution *random wear leveling*. The resulting layout is a result of both the flash allocator and the random wear leveling, but independent of the input patterns (i.e., history independence can be achieved), because: 1) the flash allocator is history independent (Sec. V-B1), and the layout resulted from it is always independent of the input patterns; 2) random wear leveling swaps blocks without relying on any other information associated with input patterns. The effectiveness of random wear leveling should be close to the start-gap wear leveling (Sec. II), as each block has an equal probability to be swapped.

SLC NAND flash is typically rated at about 100K P/E cycles [3]), and incorporating random wear leveling may be acceptable, as each block may likely be swapped before wearing out. However, modern MLC only has approximately $1K$-$10K$ P/E cycles [3]), thus incorporating random wear leveling may be problematic, as a block may never be swapped before it gets worn out.

**Wear leveling based on epoch counts.** It is unclear however whether we can do better than random wear leveling without tracking at least *some* history. Thus, instead of optimal wear leveling and long-term erasure counts that may leak too much information, we propose an epoch-based mechanism in which only a controllable amount of erasure count information is kept for a limited amount of time (the epoch). Wear leveling decisions are then only based on this limited information. Epochs are measured in time or number of transactions and may be adjusted to minimize a meaningful measure of history leak while maximizing the uniformity of the resulting wear layout and thus the device lifetime. In effect, what we are proposing is to explore the impact of reducing the amount of tracked history on the wear leveling effectiveness, and eventually balance a leakage - wear leveling effectiveness trade-off.

The hope is that even with a controlled, significantly smaller amount of historical information, acceptably effective wear leveling can still be achieved. We call this wear leveling solution *epoch wear leveling*. At the end of an epoch, the erasure counts will be re-set to 0, such that history leakage is controlled. Note that, when compared to global wear leveling, epoch wear leveling seems to be unavoidably less effective. However, the additional wear information provides significant hope that it can perform significantly better than random wear leveling.

**Handling the mappings between logical blocks and physical blocks.** We use "MAP I" to denote the map between logical blocks and physical pages. We can store MAP I in flash. However, updating MAP I will be expensive since flash is not update-friendly. A better solution is to store MAP I in RAM, as most flash devices are equipped with a certain amount of built-in RAM. For example, Jasmine OpenSSD Platform has 64GB flash and 64MB DRAM [4], Cosmos OpenSSD Platform can support up to 512GB flash and has 1GB DRAM [1], and LPC-H3131 has 500MB flash and 32MB DRAM [7]. In practice, the MAP I table is usually small in size, and is thus possible to be kept in RAM. For example, for a 64GB flash with 4KB page size and 128KB block size, MAP I is approximately $1.2MB$ in size. Storing MAP I in RAM will be advantageous, since RAM is update friendly. However, data stored in RAM will be vulnerable to system failures, due to the volatile nature of RAM and may be lost upon unexpected accidental events such as power failure.

To be able to recover MAP I after system failures, we also embed MAP I data in erase blocks. We store the logical block ID in a header of each flash block. When the flash device is gracefully turned off, MAP I is committed to flash. In the case of a power failure, the lack of this commit will be noticed upon reboot and MAP I can be reconstructed by reading the headers of the blocks in flash. For a 64GB flash device with 4KB page size and 128KB block size, this requires to read 2GB data, which can be done in 10 seconds at a 200MB/s throughput.

A simple commit flag-based mechanism can be used for failure detection. A flash-stored flag is initially reset to 0. When the flash device boots successfully, the flag will be set to 1. When the flash device is powered off normally, the flag will be re-set to 0. The flag is maintained smartly in one special erase block which can also change over time for wear leveling – after writing the address of its future location in its previous block. A number of replicas can also be maintained similar to superblock copies in file systems.

**Handling epoch counts.** Physical block erase counts for the current epoch ("epoch counts") are also maintained in a table similar to MAP I. The epoch counts table is small and can be stored in RAM. No power failure mechanism will be implemented under the assumption that power failures are rare and that keeping extremely accurate epoch counts is not critical. Upon system failures, we can simply discard the old set of epoch counts.

## VI. ANALYSIS AND DISCUSSION

### A. Security Analysis

**The basic HiFlash scheme can achieve SHI.** The history independent flash allocator (Sec. V-B1) relies on a bijective mapping between virtual block IDs and flash pages, which guarantees that (i) data being written to the same virtual block will always be assigned to the same flash page; (ii) data being written to two different virtual blocks will always be assigned to two different flash pages. Thus, the resulting layout of flash memory is always canonical, and canonical representations can achieve SHI (Sec. II), and can defend against a multiple-access adversary.

**The improved HiFlash scheme with random wear leveling can achieve WHI.** In the improved HiFlash scheme with random wear leveling, the layout is a result of both the flash allocator and the random wear leveling, but fully independent of the input patterns (Sec. V-B2). Thus, the scheme can provide at least weak history independence. Since the resulting layout is randomized by random wear leveling, two conclusions can be reached: (i) The scheme cannot provide SHI, as SHI requires canonical layout up to the initial randomness (Sec. II); (ii) The scheme can achieve WHI (i.e., can defend against the one-access adversary), as random wear leveling swaps a random (uniformly random) subset of blocks during each epoch.

**Quantify the leakage in the improved** HiFlash **scheme with epoch wear leveling.** During epoch $j$, by accessing the epoch counts, the adversary can identify an operation sequence that led to current state (from initial randomness) with a probability no larger than $\frac{\binom{m+c-1}{m-1}}{m^{cj}}$ (Appendix A), where $m$ is the total number of erase blocks in flash, and $c$ is the total number of erase operations in an epoch. This probability is very small in practice, e.g., for $m = 1000, c = 10, j = 2$, it is approximately $10^{-30}$. Thus, accessing the epoch counts provides only a small, upper-bound advantage to the adversary in identifying the exact operation sequence and input patterns that lead to the current state.

It is important however to note that this is not entirely accurate since the inability of the adversary to identify the exact sequence does not necessarily imply an inability to find out something else about the sequence. It is the subject of future work to further understand this.

*B. Discussion*

**Bad block management.** A flash device will degrade over time and develop bad blocks. Thus, we need to keep track of bad blocks and avoid using them in the future. This usually requires using a table to store the IDs of bad blocks. As this bad block table is usually small, it can be stored in RAM. When the flash device is powered off normally, the bad block table will be committed to flash. When a power failure happens, the lack of this commit (Sec. V-B2) will be noticed upon reboot and the bad block table can be reconstructed by taking a pass of the flash to check each block. Please note that bad blocks necessarily leak some historical information (i.e., that they were written often) – but since they involve hardware failure it is unclear how to hide this information from an adversary with access to the device.

**TRIM and history independence.** An ATA TRIM command [13] allows an external party to inform an SSD which blocks of data are no longer in use and can be wiped internally. Flash devices like MMC and SD provide similar functionality to the ATA TRIM [13]. Conventional file systems usually handle delete operations by flagging the deleted blocks as "unused". Thus, the underlying storage media would not know which sectors/pages can be considered as free space. However, it would be advantageous to notify the SSD when the files are deleted as the SSD will know which pages are invalid, and not to preserve the content on those pages during block erasures, wear leveling, etc. The TRIM command is designed for this notification purpose. Upon receiving a TRIM command, commodity SSDs may not wipe the invalid pages [15] immediately. This however, may conflict with history independence, as the deleted data may remain in flash. History independence requires reclaiming the invalid pages immediately, which requires performing a block erasure for each invalid page in the worse case. One optimization could be to cache TRIM commands, and perform them together (Sec. V-B1), as deletions in block devices also exhibit certain degree of locality.

### VII. IMPLEMENTATION AND SIMULATIONS

*A. Implementation*

We implemented HiFlash in an actual flash device using OpenNFM [10], an open source NAND flash controller framework. We extensively modified the OpenNFM code base, and built a History independent NAND Flash Manager (HiNFM). To be consistent with OpenNFM framework, we view the flash memory as two layers, a logical layer and a physical layer. The flash allocator can assign virtual block writes to the logical layer, while bad block management and wear leveling are independently performed between the logical and physical layer. Logical layer erase blocks are called *logical erase blocks* (LEB), while physical layer erase blocks are called *physical erase blocks* (PEB). An LEB can be viewed as a special erase block (logically exists), which has an underlying PEB, and has the following properties: 1) It will not get corrupted, as its underlying PEB will be changed to a new one if the old one is corrupted. **MAP I is used to keep track of mappings between LEBs and PEBs**. 2) It does not need to worry about wear, as its underlying PEB will be swapped by wear leveling.

*1) HiNFM Design:* Similar to OpenNFM, we adopt an architecture consisting of three layers: FTL, UBI and MTD (Figure 5). FTL mainly handles mappings between virtual blocks and LEB pages. FTL provides a uniform block device interface to an external party (Sec. III-A). UBI mainly takes care of wear leveling and bad block management. It handles mappings between LEBs and PEBs, such that: 1) it updates the mappings between LEBs and PEBs in the process of wear leveling; 2) it can re-map an LEB to a new good PEB if the old PEB is corrupted (i.e., bad block management). UBI allows the FTL to read/write flash memory without worrying about bad blocks and wear. MTD provides a raw flash abstraction. It allows the UBI to read, write and erase raw flash without being bothered by physical characteristics of different flash chips.

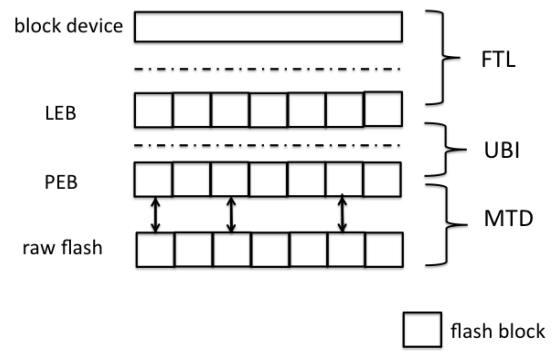

Fig. 5. HiNFM layers.

HiNFM history independence features are incorporated into the UBI and FTL layer.

**MTD.** MTD is built on top of raw flash, and mainly provides three uniform APIs to allow the UBI to read, write and erase raw flash:

- MTD_Read(PEB_index, offset, &data): read data from a PEB page identified by PEB_index and offset

- MTD_Write(PEB_index, offset, data): write data to a PEB page identified by PEB_index and offset
- MTD_Erase(PEB_index): erase the PEB identified by PEB_index

**UBI.** UBI is built on top of MTD, and uses the APIs provided by MTD to read/write PEB pages or erase PEB blocks. UBI simply views the flash as a collection of PEBs without being bothered by the underlying flash specifications. Our design for UBI is shown in Figure 6. We reserve a certain number of PEBs as free blocks (e.g., 2%), which will be used for replacing bad blocks. We also reserve a small number of PEBs (e.g., 0.1%) for storing the bad block table (Sec. VI-B) and MAP I (Sec. V-B2). We call these PEBs "anchor blocks". As the bad block table and MAP I are both small in size, they can be stored and updated in RAM. They will only be stored in the anchor blocks when the flash device is shut down normally, which will not add too much wear to the anchor blocks. However, to avoid being worn out, the anchor blocks will be periodically swapped with a new set of PEBs. UBI mainly provides three uniform APIs to allow FTL to read/write LEB pages or erase an LEB:

- UBI_Read(LEB_index, offset, &data): read data from an LEB page identified by LEB_index and offset
- UBI_Write(LEB_index, offset, data): write data to an LEB page identified by LEB_index and offset
- UBI_Erase(LEB_index): erase an LEB identified by LEB_index, which will cause an erasure over the corresponding PEB

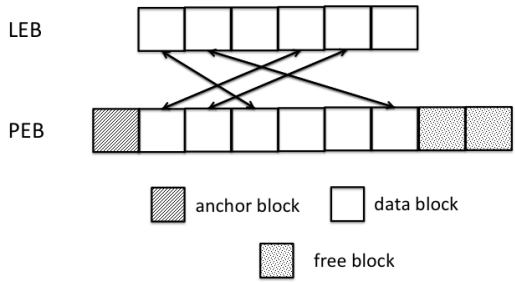

Fig. 6. UBI remapping in HiNFM.

**FTL.** FTL is built on top of UBI, and uses the APIs provided by UBI to read/write LEB pages or erase LEBs. FTL simply views the flash as a collection of LEBs, without having to handle bad blocks and wear leveling. The HiFlash FTL design reserves a certain number of LEBs for journal blocks (e.g., 1), and the remaining LEBs are used as regular data blocks. Note that LEBs will never wear out as the underlying UBI will take care of wear leveling (i.e., if an LEB generates too much wear to the corresponding PEB, it will be relocated to another PEB by wear leveling). The flash allocator (Sec. V-B1) assigns virtual block writes to LEB pages. The writes are first cached in the journal blocks. When the journal blocks are filled, we will perform all the writes together. The journal LEBs are then erased and re-used.

*2) Implementation and Experimental Results:* We ported HiNFM to LPC-H3131 [7], a development board equipped with 180 MHz ARM microcontroller, 500MB NAND flash and 32 MB SDRAM. The flash has 128-KB blocks and 2KB pages – each erase block is composed of 64 pages. The resulting board can be used as a history independent USB 2.0 flash drive. As LPC-H3131 has approximately 4000 erase blocks in its NAND flash, the MAP I table will be approximately 6KB in size (MAP I will contain 4000 mappings, each of which can be represented by 12 bits). Similarly, the bad block table will be also approximately $6KB$ in size.

We benchmarked HiNFM using fio [2] (when running fio, we used non-buffered io), running in a host computer with 8 Intel i7 CPUs at 1.60GHz, 10GB RAM and Windows 8 Pro 64-bit. For comparison, we also benchmarked OpenNFM. As the latest version of OpenNFM does not incorporate wear leveling, we added global wear leveling support to it as well as some other optimizations (e.g., storing the bad block table in RAM), and **used this modified version of OpenNFM for comparison**.

**Implementation and deployment issues.** We note that LPC-H3131 is not compatible with USB 3.0. We also observed that the MTD_Check function in the OpenNFM framework is not reliable, and cannot guarantee always returning a correct result. Thus, using this function to check whether a block is bad will not be reliable.

**Experimental results.** For both HiNFM and OpenNFM, we choose the same wear leveling epoch (e.g., 100 erase operations), and the same number of blocks being swapped (e.g., 1) during each epoch. In HiNFM, to mitigate write amplification, we did the following according to Sec. V-B1: 1) we increase the virtual block size; 2) we used 1 journal block (which is an LEB block an will not wear out as UBI takes care of wear leveling) to cache writes.

Benchmarking results for both HiNFM and OpenNFM are shown in Figure 7. We have several observations:

(a) The sequential read throughput of HiNFM is close to that of OpenNFM ($0.9\times$, as shown in Figure 7(a)). Random read of HiNFM has approximately $1.2\times$ throughput compared to OpenNFM (Figure 7(a)). This is because, HiNFM FTL can efficiently locate an LEB page by simply performing the F function (Sec. III-A) over the virtual block ID. In contrast, OpenNFM FTL searches through a flash-stored map table keeping track of the mappings between virtual block IDs and LEB pages – this is necessary since virtual block IDs and LEB pages do not have a fixed relation in OpenNFM. The map is stored in flash and only a small portion is cached in RAM, as it keeps track of pages and will be large in size. However, sequential read in OpenNFM is fast as the cached mappings preserve locality of virtual blocks.

(b) For sequential write of HiNFM, we observed approximately $4\times$ degradation in throughput compared to OpenNFM (Figure 7(b)-7(d)). For random write of HiNFM, we observed more degradation in throughput, $5\times$ when the virtual block size is increased to 64KB (Figure 7(d)). The write performance degradation is mainly due to write amplification. OpenNFM does not provide history independence and thus can always write the data to new empty pages, resulting in significantly lower write amplification.

For HiNFM however, write amplification may be the unavoidable price to pay because we need to perform in-place update to achieve history independence, and in-place update is expensive for flash memory. Utilizing temporal locality and increasing the virtual block size can help to mitigate write amplification, but unfortunately cannot eliminate it. Note that, when the virtual block size is small, sequential write in HiNFM seems to be much faster than random write (Figure 7(b)). This is because, caching writes can help improve the sequential write performance as writes performed sequentially usually exhibit good locality. However, writes performed randomly usually do not exhibit locality due to randomness. Increasing virtual block size can help improve the performance of random write (Figure 7(b)-7(d)), as it can help reduce the gap between the unit of an in-place update and the unit of an erasure (i.e., one erase block), mitigating write amplification (Sec. II).

*B. Simulations*

To fine-tune our design we performed extensive simulations using real-life workloads. We evaluated the effectiveness of utilizing temporal locality to mitigate write amplification; we measured the distribution of block erasures over flash when incorporating different wear leveling mechanisms into HiFlash. Each erase block consists of 64 4-KB pages.

We used three representative workloads (Table I), selected from the 1-week block I/O traces of enterprise servers at Microsoft Research Cambridge [47]. Workload data (Table I shows that small-sized (about 4KB) writes dominate.

| Workload name | web_1 | wdev_0 | hm_0 |
|---|---|---|---|
| Workload type | web/SQL server | test web server | monitoring server |
| # of total writes | 73,833 | 913,732 | 4,060,610 |
| Workload category | small size | medium size | large size |

TABLE I
SUMMARY OF OUR WORKLOADS: WE CHOOSE A SMALL-SIZE, A MEDIUM-SIZE AND A LARGE-SIZE WORKLOAD, RESPECTIVELY.

|  | web_1 | wdev_0 | hm_0 |
|---|---|---|---|
| 0.5KB | 16.7% | 4.7% | 14.1% |
| 4KB | 59.6% | 61.8% | 48.4% |
| 8KB | 7.6% | 6.6% | 10.7% |
| 16KB | 1.0% | 11.1% | 5.9% |
| Others | 15.1% | 15.8% | 20.9% |

TABLE II
WRITE SIZES FOR DIFFERENT WORKLOADS.

**Mitigating write amplification.** We simulated the effectiveness of utilizing temporal locality to mitigate write amplification. For different workloads, we measured the total number of writes and erasures performed on the flash by varying the number of cached writes from 0 to 256. The simulation results are shown in Figure 8(a) and 8(b). We observed that, (i) Compared to no caching, the total number of writes and erasures can be reduced by approximately an order of magnitude by caching 64 writes, and (ii) overall, as expected, the total number of writes and erasures decreases with increasing number of cached writes. This justifies the effectiveness of using temporal locality to mitigate write amplification for real workloads.

**Evaluating different wear leveling mechanisms.** In this simulation, we incorporated different wear leveling solutions into HiFlash, and evaluated how they affect the distribution of block erasures. A wear leveling mechanism is said to be effective if the erasures are distributed evenly across flash blocks. We evaluated the effectiveness by relying on the maximum erasure count in a distribution. We perform two sets of simulations:

- We fix the epoch size as 100 erase operations, and swap 10 erase blocks at the end of each epoch. Simulation results for various workloads are shown in Figure 9 (no wear leveling), Figure 10 (random wear leveling), Figure 11 (epoch wear leveling) and Figure 12 (global wear leveling [23]).

- We vary the epoch size (120, 140, 160 and 180 erase operations) to study how epoch size will influence the distribution of block erasures when incorporating epoch wear leveling. Correspondingly, we swap 12, 14, 16 and 18 blocks at the end of each epoch respectively. This can ensure that the overhead brought by swapping blocks remains the same for the aforementioned four cases. Simulation results for workload "hm_0" are shown in Figure 13 (as we have similar observations for different workloads, we only show the set of simulation results for the large-size workload).

We note several observations: (i) Lack of wear leveling causes significant write unevenness (Figure 9), as the basic HiFlash scheme bijectively maps virtual blocks to flash pages, passing the write unevenness on the block device to flash. (ii) Compared to no wear leveling, random wear leveling can improve the effectiveness by approximately an order of magnitude (Figure 10). (iii) Compared to random wear leveling, epoch wear leveling is approximately an order of magnitude more effective (Figure 11). However, its effectiveness is only about 50% of that of full-fledged global wear leveling (Figure 12). Epoch wear leveling seems to have the potential to properly balance a leakage - wear leveling effectiveness trade-off. (iv) Increasing epoch size will improve the effectiveness of epoch wear leveling (Figure 13). This is because: when increasing the epoch size, more history information can be relied on to balance wear at the end of each epoch. Note that increasing epoch size will lead to more leakage in each epoch, as the adversary can learn the epoch counts when having access to the flash (Sec. V-B2).

## VIII. RELATED WORK

**History Independence.** Micciancio et al. [42] initiated the research of history independence. They designed an oblivious 2-3 tree with the property that the topology of the tree does not leak the sequence of operations that led to it. Several choices (e.g., whether a non-leaf node will have two or three child nodes is randomized, re-balancing during local modifications) are randomized. This enables the tree to have a probability distribution of nodes that is independent of the sequence of operations. Naor et al. [46] were the first to introduce the notions of weak and strong history independence (WHI and SHI). They designed a SHI hash table that supports search and insert operations. The hash table construction is similar to linear probing [41] except for the collision resolution. Golovin et al. [20, 29] designed a history independent hash table that can support search, insert and delete operations, based on the stable matching property of the *Gale-Shapley Stable Marriage* [27] algorithm. Some other history independent data structures have been designed, including Cuckoo Hashing [45], B-Treaps [30], B-SkipList [31], R-Trees [54].

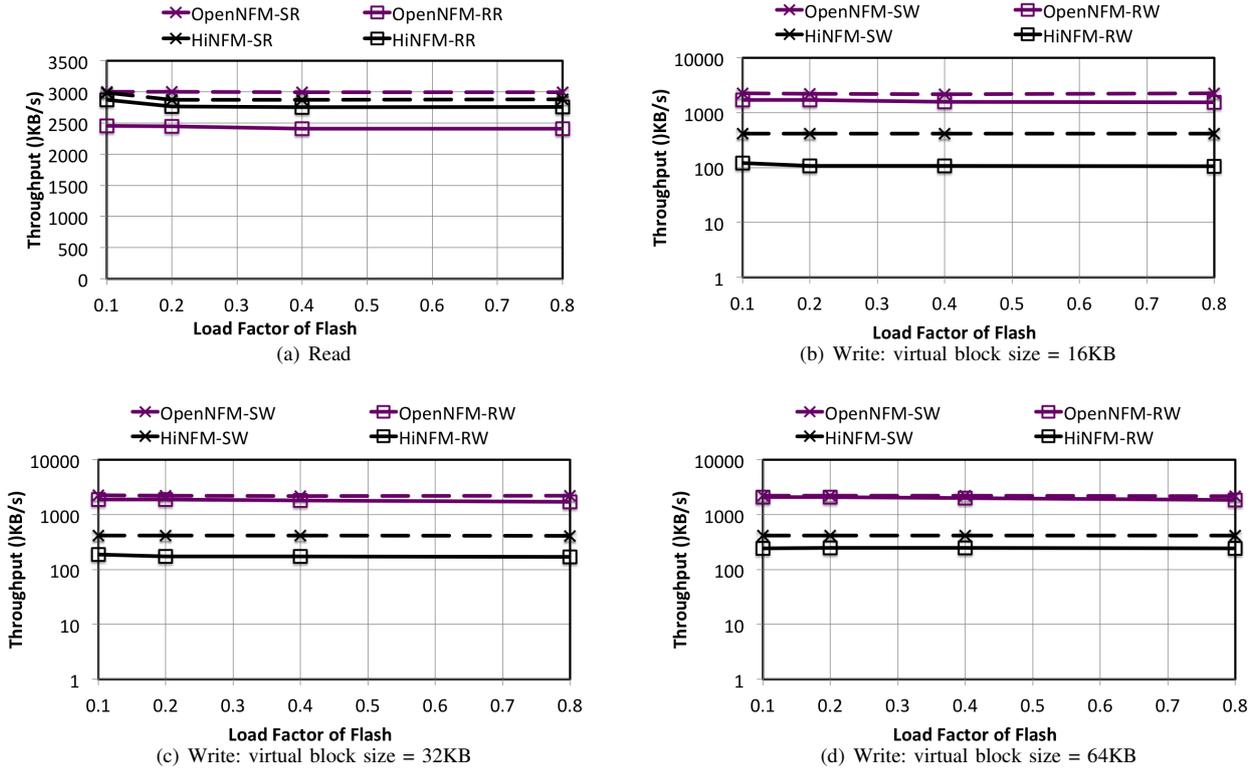

Fig. 7. Read/Write throughput of OpenNFM and HiNFM: SR - sequential read, RR - random read, SW - sequential write, RW - random write.

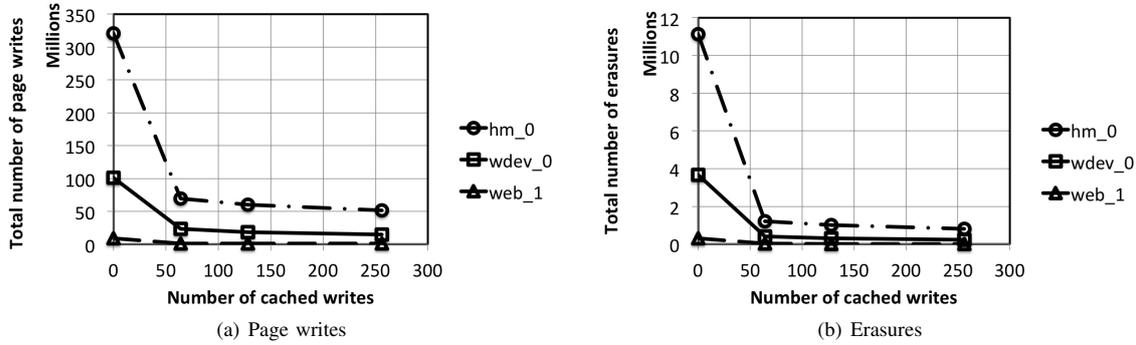

Fig. 8. Mitigating write amplification by caching multiple writes: simulation results based on real workloads.

Hartline et al. [32, 33] showed that strong history independence necessarily implies that the data structure has canonical representations up to initial randomness. Buchbinder et al. [21] obtained the first time complexity separation between the weak and the strong notions of history independent data structures (e.g., for both heap and queue).

**History independence on write once storage.** Molnar et al. [43] designed history independence schemes for write-once storage, e.g., in voting machines, such that the contents of the voting record will not reveal the order in which ballots were cast. They proposed multiple candidate constructions for how to organize storage of write-in votes. The copyover list construction requires $O(n^2)$ space to store $n$ keys. The single pooled lexicographic chain table requires $O(nlog^2 n)$ space. The most space-efficient solution is the single pooled random placement table, in which new elements are inserted at random locations on the write-once storage. Although simple and space-efficient, the random approach requires random bits to be hidden from the adversary. Moran et al. [44] proposed a solution that requires $O(n \cdot polylog(N))$ space, to store a set of at most $n$ keys from a large universe of size $N$.

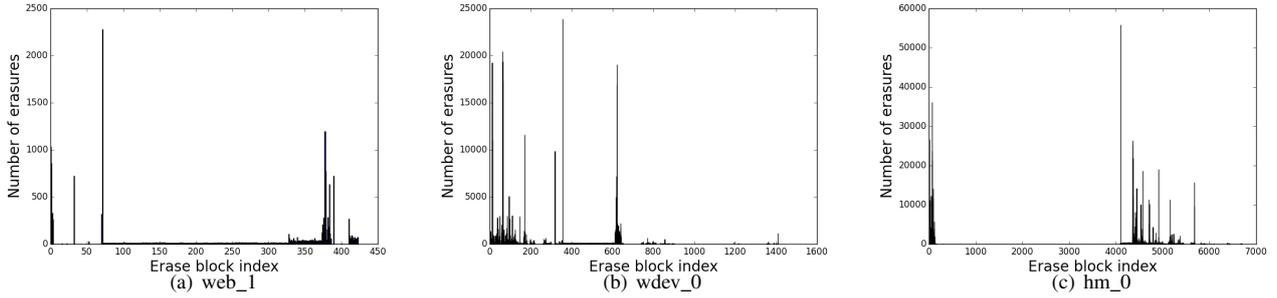

Fig. 9. The distribution of block erasures for various workloads in the basic HiFlash: **no wear leveling**.

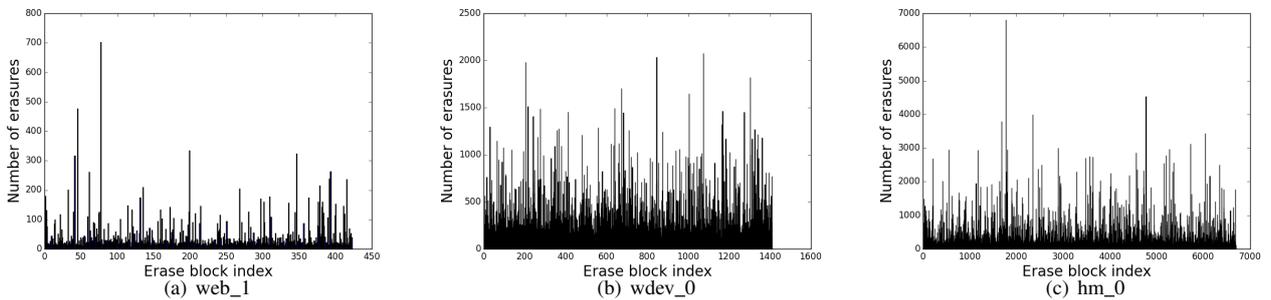

Fig. 10. The distribution of block erasures for various workloads in the improved HiFlash: using **random wear leveling**; the epoch size is 100 erase operations.

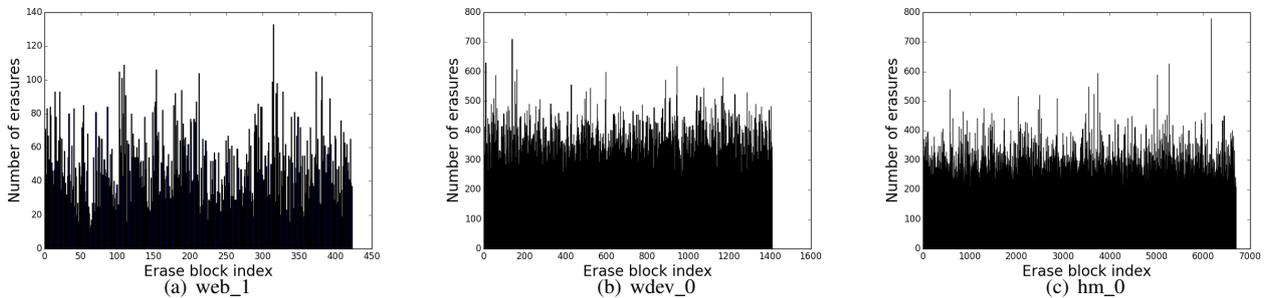

Fig. 11. The distribution of block erasures for various workloads in the improved HiFlash: using **epoch wear leveling**; epoch size is 100 erase operations.

**Secure deletion on solid state drives.** Sun et al. [52] identified zero overwriting and block cleaning, as techniques to securely delete data on flash storage. They also proposed a hybrid scheme that adaptively applies the more efficient solution. Lee et al. [37, 38] proposed an encryption-based secure deletion scheme for YAFFS, a log-structured file system. The current and the previous keys of a file are forced to be stored in the same block of the flash memory. Thus, a file can be deleted by a single block erase. Lee et al. [36] extended this solution with standard data sanitization operations on the key containing blocks.

Reardon et al. [50] introduced three techniques for secure deletion in YAFFS: purging and ballooning at the user-level, and zero overwriting at the kernel level. In [49], they introduced the Data Node Encrypted File System (DNEFS), which can achieve secure deletion against a computationally-bounded single access adversary. DNEFS encrypts each data node with a different key and collocates the keys in a key storage area on the flash. They instantiated DNEFS for UBIFS, and built UBIFSec, which can achieve fine-grained deletion and provide a guaranteed upper bound on deletion latency.

Swanson et al. [53] combined encryption and erasure based methods to achieve secure deletion. The combination can provide almost instant erasure along with verifiability. By observing that programming individual pages is possible, Wei et al. [55] proposed to use scrubbing to efficiently sanitize flash pages. They presented and evaluated three different scrubbing

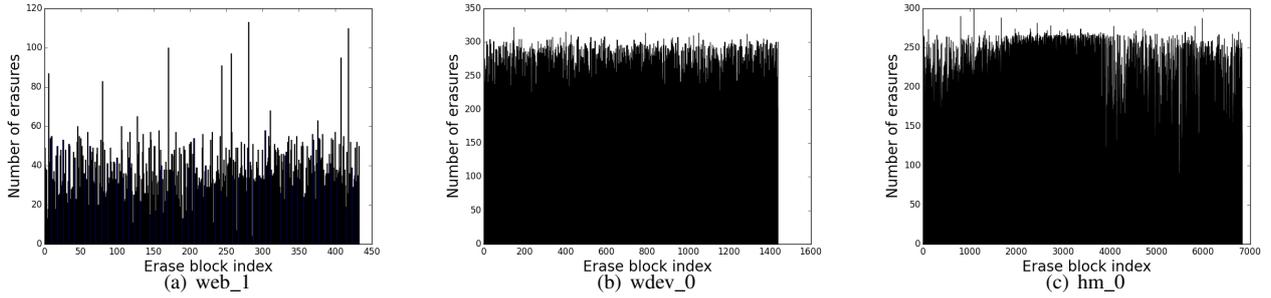

Fig. 12. The distribution of block erasures for various workloads: using **global wear leveling**; the epoch size is 100 erase operations.

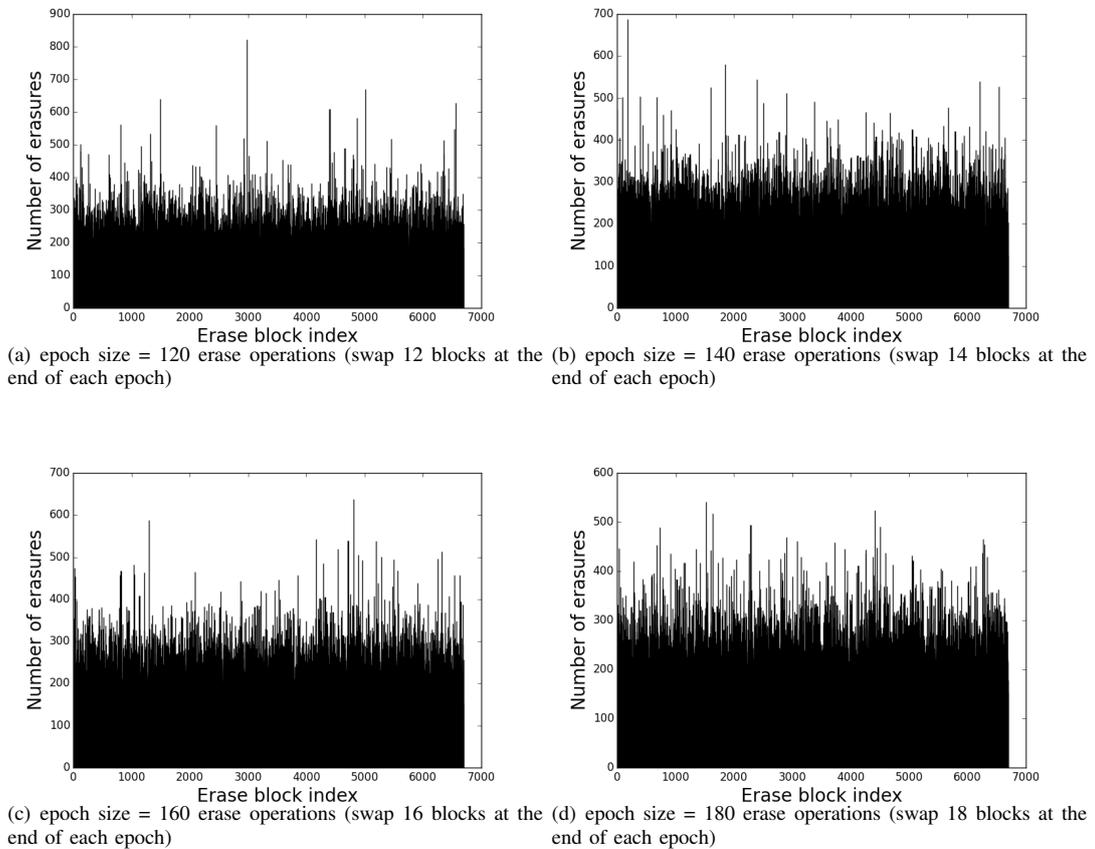

Fig. 13. The distribution of block erasures for epoch wear leveling under different epoch sizes.

methods that make different trade-offs between performance and security.

Diesburg et al. [26] proposed TrueErase, a framework that deletes data and metadata upon user request. TrueErase correctly propagated secure deletion information all the way from the user to the storage. To securely delete a page from a flash block, they copy other in-use pages from the current flash block to other areas, and mark those pages as unused in the block. The page to be deleted is then marked invalid, and the current flash block can be cleared via a block erasure.

To securely remove data from flash, the aforementioned studies used encryption [36]–[38, 49], block erasure [26, 52, 53] or scrubbing [52, 55]. However, none of them can provide history independent flash layout. In all of them, the write history of the data would implicitly leave artifacts in the layout of flash, which may be used as an oracle by the attacker to infer the past existence of deleted data. HiFlash however, can eliminate these history-related oracles from flash layout, making it impossible to recover the deleted data from flash devices.

**Other related work.** Caching is pervasively used in systems (e.g., Kells [22]) to improve performance. Caching in HiFlash is slightly different in that it does not rely on volatile RAM for caching, to mitigate power failure issues. Instead, we use the non-volatile flash for caching (Sec. VII-A2). The writes to flash will be first stored in journal blocks sequentially. After a certain number of writes are accumulated, the writes are performed to the actual flash locations (the total number of erasures required is expected to be reduced as the cached writes will belong to fewer different erase blocks).

## IX. CONCLUSION

In this paper, we design HiFlash, a first history independent flash device. HiFlash achieves history independence by building around a bijective mapping between virtual block IDs and flash pages, optimizing write amplification by utilizing temporal locality, and a history independence friendly wear leveling mechanism that allows HiFlash to smartly and advantageously trade off a tunable small amount of history leakage for a significant increase in the device's lifetime. A prototype implemented in an actual flash device and extensive simulations validate the effectiveness of HiFlash.

## APPENDIX

Assume the flash consists of $m$ erase blocks. An epoch is chosen as $c$ erase operations. During epoch $j$ (where $j \geq 1$), the adversary accesses the flash device, and obtains the entire epoch counts. We aim to quantify the probability that the adversary can learn the operation sequence which led to current layout (from the initial randomness). We use $P_j$ to denote this probability. To simplify this problem, we assume the writes are performed on erase blocks. Equivalently, an epoch has $c$ writes. In the following, we start with some basic cases, and then generalize the results.

**Case 1**: $m = 2, c = 2$. We use A and B to identify the two erase blocks respectively. The adversary accesses the flash at the end of epoch $j$, and tries to derive the write sequence starting from the initial randomness (i.e., the beginning of epoch 1).
(1) If the adversary accesses the flash at the end of epoch 1:

The possible distributions of epoch counts on block A and B are (2, 0), (1, 1), (0, 2). The adversary may observe one of these three distributions, and may identify the write sequence as follows:

(2, 0): the write sequence can only be A, A;

(0, 2): the write sequence can only be B, B;

(1, 1): the write sequence can be either A, B or B, A.

Thus, by observing the epoch counts at the end of epoch 1, $P_1$ will be $\frac{1}{4} \times 1 + \frac{1}{4} \times 1 + \frac{1}{2} \times \frac{1}{2} = \frac{3}{4} = \frac{3}{(2^2)^1}$.
(2) If the adversary accesses the flash at the end of epoch 2:

As the adversary has no knowledge about epoch 1, it can only guess the write sequence in epoch 1. Each possible write sequence in epoch 1 has an equal probability of $\frac{1}{4}$, as there are 4 possible write sequences. Thus, $P_2$ can be calculated as $\frac{1}{4} \times (\frac{1}{4} \times 1 \times + \frac{1}{4} \times 1 + \frac{1}{2} \times \frac{1}{2}) = \frac{3}{4^2} = \frac{3}{(2^2)^2}$.
(3) If the adversary accesses the flash at the end of epoch $j$:

We simply generalize $P_j$ as $\frac{3}{4^j} = \frac{3}{(2^2)^j}$, in which 3 is the total number of distributions of epoch counts.

**Case 2**: m=3, c=2. We use A, B and C to identify the 3 erase blocks respectively. Similarly, the adversary accesses the flash at the end of epoch $j$, and tries to derive the write sequence starting from the initial randomness.
(1) If the adversary accesses the flash at the end of epoch 1:

The possible distributions of epoch counts on these three erase blocks are (2, 0, 0), (0, 2, 0), (0, 0, 2), (1, 1, 0), (1, 0, 1), (0, 1, 1). The adversary will observe one of the six distributions, and identify the write sequence as follows:

(2, 0, 0): the write sequence can only be A, A;

(0, 2, 0): the write sequence can only be B, B;

(0, 0, 2): the write sequence can only be C, C;

(1, 1, 0): the write sequence can be either A, B or B, A;

(1, 0, 1): the write sequence can be either A, C or C, A;

(0, 1, 1): the write sequence can be either B, C or C, B.

Thus, $P_1 = \frac{1}{9} \times 1 + \frac{1}{9} \times 1 + \frac{1}{9} \times 1 + \frac{2}{9} \times \frac{1}{2} + \frac{2}{9} \times \frac{1}{2} + \frac{2}{9} \times \frac{1}{2} = \frac{6}{9} = \frac{6}{(3^2)^1}$.
(2) If the adversary accesses the flash at the end of epoch 2:

The adversary has no knowledge on epoch 1, and can only guess the write sequence in epoch 1 with probability $\frac{1}{9}$, as there are 9 possible write sequences in an epoch. Thus, $P_2 = \frac{1}{9} \times (\frac{1}{9} \times 1 + \frac{1}{9} \times 1 + \frac{1}{9} \times 1 + \frac{2}{9} \times \frac{1}{2} + \frac{2}{9} \times \frac{1}{2} + \frac{2}{9} \times \frac{1}{2}) = \frac{6}{9^2} = \frac{6}{(3^2)^2}$.
(3) If the adversary accesses the flash at the end of epoch $j$:

$P_j = \frac{6}{9^j} = \frac{6}{(3^2)^j}$, in which 6 is the total number of distributions of epoch counts.

**General case**:

Let $\alpha$ be the total number of distributions of epoch counts, e.g., $\alpha = 3$ for the case "$m = 2, c = 2$", and $\alpha = 6$ for the case "$m = 3, c = 2$". For general values of $m$ and $c$, $P_j = \frac{\alpha}{m^{cj}}$.

We further quantify $\alpha$. By distributing $c$ writes to $m$ blocks, we would like to quantify the total number of possible distributions. This is actually a stars and bars problem [12]. The answer is $\binom{m+c-1}{m-1}$, i.e., $\alpha = \binom{m+c-1}{m-1}$.